\documentclass[12pt]{article}
\pdfoutput=1
\usepackage[numbers,sort&compress]{natbib}
\usepackage{graphicx}
\usepackage{amsmath,graphicx,amssymb,subfigure}
\usepackage[usenames,dvipsnames]{color}
\usepackage{amssymb}
\usepackage{amsmath}
\usepackage{isomath}
\usepackage{amsthm}
\usepackage{enumerate}
\usepackage{stmaryrd}
\usepackage{amsfonts}
\usepackage[all]{xy}
\usepackage[section]{placeins}
\usepackage[colorlinks=true,linktocpage=true,linkcolor=blue,citecolor=blue]{hyperref}
\newcommand{\mt}[1]{\textrm{\tiny #1}}
\newcommand{\be}{\begin{equation}}
\newcommand{\ee}{\end{equation}}
\newcommand{\bea}{\begin{eqnarray}}
\newcommand{\eea}{\end{eqnarray}}
\newcommand{\rh}{r_\mt{H}}

\begin{document}

\bibliographystyle{hieeetr}

\pagestyle{plain}
\setcounter{page}{1}

\begin{titlepage}

\begin{center}

\vskip 30mm

{\Large\bf Wilsonian RG flow approach to holographic transport with momentum dissipation}

\vskip 0.8cm

{\bf Yu Tian}$^{1,4}$,~~~ {\bf Xian-Hui Ge}$^{2,3,4}$,~~~ {\bf Shao-Feng Wu}$^{2,4,5}$\\
$^1${\it \small School of Physics, University of Chinese Academy of Sciences, Beijing, 100049, P.R. China}\\
$^2${\it \small Department of Physics, Shanghai University, Shanghai 200444, P.R. China}\\
$^3${\it\small Department of Physics, University of California at San Diego, CA92093, USA}\\
$^4${\it\small Shanghai Key Laboratory of High Temperature Superconductors, Shanghai 200444, P.R. China}\\
$^5${\it \small Shanghai Key Lab for Astrophysics, 100 Guilin Road, 200234 Shanghai, P.R. China}\\
%{\sf{gexh@shu.edu.cn}}, {\sf{ytian@ucas.ac.cn}}, {\sf{sfwu@shu.edu.cn}}
\medskip

\vspace{5mm}
\vspace{5mm}

\begin{abstract}
We systematically present a new approach for studying the coupled linear transport of holographic systems. In this approach, the set of equations for the linear perturbations can be reduced to a first-order nonlinear ordinary differential equation expressed as the radial (renormalization group) flow equation of the transport matrices. As an important application, we use this approach to compute the DC and AC conductivities of a holographic model with momentum dissipation, which can be easily read off from the nonlinear flow equations. This method also works for transport in the presence of an external magnetic field.
\end{abstract}
\end{center}
\noindent
\end{titlepage}

\section{Introduction}
The AdS/CFT correspondence provides a powerful tool in studying strongly coupled systems in condensed matter physics. Recently, great efforts have been devoted to exploring translational
symmetry broken systems holographically, aimed to mimic real condensed systems \cite{qlattice,john,lingprl,withers,kim,Gouteraux15,matteo,GLNS,amoretti,amoretti1,gautlett1,BP,bkp,gautlett2,gautlett3,blaise2,kw,WK15,lv16,pengliu,blaise3,schalm,gsw2015,cheng,mb1,mb2}. In the AdS/CFT correspondence one can move the boundary inwards by exploiting a holographic version of the Wilsonian renormalization group (RG). The emergence of the holographic direction is related to some coarse graining procedure. The essential idea of holographic Wilsonian Renormalization Group  approach is to integrate out the bulk field from the boundary up to some intermediate radial distance.
The radial direction in the bulk marks the energy scale of the boundary theory and the radial flow in the bulk geometry can be interpreted as the renormalization group flow of
the boundary theory.

In translational symmetric systems, the holographic Wilsonian RG flow approach has been widely studied. It was proved in \cite{sin} that several approaches to RG flow of transport coefficients are equivalent: They are sliding membrane paradigm \cite{liu1,liu}, Wilsonian fluid/gravity \cite{strominger} and Holographic Wilsonian RG \cite{po,sken,ge}. For charged black holes, metric fluctuations and Maxwell fluctuations will mix. Such mixing effect of metric and Maxwell fluctuations in charged black hole background is important in that it renders the transverse vector modes of Maxwell field diffusive. In \cite{sin2}, the authors wrote down the coupled flow equations for the ``electric conductivity", ``momentum current conductivity" and mixing parameters defined there (see equations (71-73) in \cite{sin2}), but have not obtained a neat flow equation in matrix form and solved the flow equations. As another drawback, those quantities defined in their way did not have the direct meaning of transport coefficients in the sense of Onsager's phenomenological relations. Actually, they in turn relied on the second order master field differential equations. In this paper, we will systematically show that this dilemma can be avoided by directly writing down the flow equation of the matrix of transport coefficients, which has an elegant form and can be regarded as the RG flow of Onsager's phenomenological relations. This approach has been preliminarily used in \cite{ge16,ge1606}.  It turns out that in the canonical case our flow equation takes the form of the matrix Riccati equation, which also appears in optimization theory and other disciplines (see, for example, \cite{optimization}).

On the other hand, up to now, there are few papers working on the translational symmetry broken system by using the Wilsonian RG flow approach, except in \cite{ge16,ge1606}. In this paper, we provide a systematic scheme on how to generalize the Wilsonian RG flow approach to coupled transport systems in the presence of momentum relaxation. We will show that one of the advantages of this method is that it effectively transforms the equations of motion from second order ordinary differential equations of linear perturbations to first order non-linear ordinary differential equations of RG flows. This in turn will greatly simplify the numerical computation of the AC conductivities.

The structure of this paper is organized as follows. In section 2, we present the general formalism for the transport coefficients from Onsager's theory. The RG flow equation from an effective bulk action is also derived. We present the bulk theory and the black hole solution in the presence of momentum relaxation in section 3. The black hole solution utilized here was previously derived by Andrade and Withers in \cite{withers} (see also \cite{mb1,mb2} for momentum relaxation in massive gravity theory). As an illustration of the efficiency of the RG flow approach, we calculate the DC
electrical conductivity, thermal conductivity and thermoelectric conductivity in terms of
the horizon data in section 3.1 and 3.2. The AC electric conductivity will be computed numerically in section 3.3. In section 3.4, we give DC transport in the presence of finite magnetic field. Discussions and conclusions are presented in section 4. In the Appendix, we provide a derivation of the flow equation in the Einstein-Maxwell theory using our general formalism.

\section{{The general formalism of the coupled flow equations}}
In this section, we give the general arguments of the formalism for  coupled flow equations of transport. We mainly follow the ideas of Onsager's phenomenological theory and extend it to the holographic situations.
\subsection{Transport coefficients for coupled transport processes}
According to Onsager, the coupled transport processes in the linear
regime are described by the following phenomenological relation:
\begin{equation}
J=\tau X,\label{eq:relation}
\end{equation}
where $X$ is the column vector formed by the sources, $J$ that formed
by the responses, and $\tau$ the matrix of the transport coefficients.
The entropy production rate is given by
\[
\Sigma=X^{T}J=X^{T}\tau X,
\]
provided $J$ is canonically conjugate to $X$.

In some cases, for example in the holographic calculation of $\tau$,
we need to express $\tau$ in terms of $X$ and $J$. Naively we
can write
\[
\tau=J/X,
\]
but that is only well-defined when $X$ (and also $J$) is just a
number, i.e. a column vector with a single component. Nevertheless,
a generalization of that expression for a multi-component $X$, i.e.
for a genuine coupled transport process, is still possible. Take the
two-component case as the simplest example. We have
\[
\left(\begin{array}{c}
J_{1}\\
J_{2}
\end{array}\right)=\tau\left(\begin{array}{c}
X_{1}\\
X_{2}
\end{array}\right),
\]
where $\tau$ is a $2\times2$ matrix. In order to ``divide''
both sides of the above equation by $X$, the only way out is to make
$X$ a square matrix. Actually, we may simply use the above equation
twice:
\[
\left(\begin{array}{c}
J_{1}^{(1)}\\
J_{2}^{(1)}
\end{array}\right)=\tau\left(\begin{array}{c}
X_{1}^{(1)}\\
X_{2}^{(1)}
\end{array}\right),\qquad\left(\begin{array}{c}
J_{1}^{(2)}\\
J_{2}^{(2)}
\end{array}\right)=\tau\left(\begin{array}{c}
X_{1}^{(2)}\\
X_{2}^{(2)}
\end{array}\right),
\]
with $X^{(1)}$ and $X^{(2)}$ two linearly independent source vectors,
which immediately means
\begin{eqnarray}
\left(\begin{array}{cc}
J_{1}^{(1)} & J_{1}^{(2)}\\
J_{2}^{(1)} & J_{2}^{(2)}
\end{array}\right)=\tau\left(\begin{array}{cc}
X_{1}^{(1)} & X_{1}^{(2)}\\
X_{2}^{(1)} & X_{2}^{(2)}
\end{array}\right).\label{eq:Onsager}
\end{eqnarray}
Due to the linear independence of $X^{(1)}$ and $X^{(2)}$, a direct
consequence of the above form is that the ``extended'' matrix $X$
on the right hand side can be inverted now, which leads to
\begin{equation}
\tau=\left(\begin{array}{cc}
J_{1}^{(1)} & J_{1}^{(2)}\\
J_{2}^{(1)} & J_{2}^{(2)}
\end{array}\right)\left(\begin{array}{cc}
X_{1}^{(1)} & X_{1}^{(2)}\\
X_{2}^{(1)} & X_{2}^{(2)}
\end{array}\right)^{-1}.\label{eq:sigma}
\end{equation}
That is the key fact on which our whole discussion is based.

For notational briefness, hereafter we will denote (\ref{eq:Onsager})
and (\ref{eq:sigma}) as
\[
\left\llbracket \begin{array}{c}
J_{1}\\
J_{2}
\end{array}\right\rrbracket =\tau\left\llbracket \begin{array}{c}
X_{1}\\
X_{2}
\end{array}\right\rrbracket,
\]
and
\[
\tau=\left\llbracket \begin{array}{c}
J_{1}\\
J_{2}
\end{array}\right\rrbracket \left\llbracket \begin{array}{c}
X_{1}\\
X_{2}
\end{array}\right\rrbracket ^{-1},
\]
respectively, where the special notation $\llbracket...\rrbracket $ should be considered as a square matrix which is introduced
for convenience.  An explicit example  for the notation  $\llbracket...\rrbracket $ is given as follows
\bea\left \llbracket \begin{array}{c}
 X_{1}\\
 X_{2}
\end{array}\right \rrbracket\equiv
 \left(\begin{array}{cc}
 X^{{(1)}}_{1} & X^{(2)}_{1}\\
 X^{{(1)}}_{2} & X^{(2)}_{2}
\end{array}\right),
\eea
where the auxiliary notations $X^{(i)}_{1}$ and $X^{(i)}_{2}$ are linearly independent sources, introduced to guarantee the source term invertible. After inverting the components in $\llbracket ...\rrbracket$, $X^{(i)}_{1}$ and $X^{(i)}_{2}$ will be not important in further calculations and it is better for us to hide them in $\llbracket ...\rrbracket$.

\subsection{{Matrix flow equation of transport coefficients in holography}}
Linear transport processes on the boundary are holographically described
by linear perturbations in the bulk. Suppose that we have an effective
bulk action
\[
I[\Phi]=\int L{[\Phi]}dr,
\]
for linear perturbations around a background bulk solution, which
corresponds to an equilibrium state on the boundary. By definition, $I$ is quadratic in the bulk
perturbation fields, which are collectively denoted as $\Phi$. Here we {assume} that the background bulk solution is homogeneous in {the time direction $t$}, so that we {can consider a single Fourier mode $e^{-i\omega t}$ of $\Phi$}. The
Euler-Lagrange equation is
\begin{equation}
\frac{d}{dr}\frac{{\delta} L}{{\delta}\Phi^{\prime}}=\frac{{\delta} L}{{\delta}\Phi}.\label{eq:Euler}
\end{equation}
{Here, as well as throughout this paper, the prime denotes the derivative with respect to $r$.} By holographic dictionary, we have
\[
X=-\partial_{t}\bar{\Phi}=i\omega\bar{\Phi},\qquad J=\frac{{\delta I^{\mathrm{os}}[\bar{\Phi}]}}{{\delta}\bar{\Phi}},
\]
for the $\omega$ mode, where $\bar{\Phi}$ is the boundary value
of $\Phi$ and $I^{\mathrm{os}}{[\bar{\Phi}]}$ is the on-shell counterpart
of the bulk action $I[\Phi]$. So Onsager's phenomenological relation
(\ref{eq:relation}) reads
\[
\frac{{\delta I^{\mathrm{os}}[\bar{\Phi}]}}{{\delta}\bar{\Phi}}=i\omega\tau\bar{\Phi},
\]
where we should bear in mind that $\bar{\Phi}$ (as well as ${\delta I^{\mathrm{os}}[\bar{\Phi}]}/{\delta}\bar{\Phi}$)
is a multi-component column vector. The flow equation of $\tau$
is obtained by derivative of the above equation with respect to $r$,
which reads
\begin{equation}
\frac{d}{dr}\frac{{\delta I^{\mathrm{os}}[\bar{\Phi}]}}{{\delta}\bar{\Phi}}=i\omega\tau^{\prime}\bar{\Phi}+i\omega\tau\bar{\Phi}^{\prime}.\label{flow1}
\end{equation}
The Hamilton-Jacobi equation tells us
\begin{equation}
\frac{{\delta I^{\mathrm{os}}[\bar{\Phi}]}}{{\delta}\bar{\Phi}}=\frac{{\delta} L}{{\delta}\Phi^{\prime}},\label{eq:Hamilton}
\end{equation}
so (\ref{flow1}) can be turned into
\[
\frac{d}{dr}\frac{{\delta} L}{{\delta}\Phi^{\prime}}=i\omega\tau^{\prime}\bar{\Phi}+i\omega\tau\bar{\Phi}^{\prime}.
\]
Upon using (\ref{eq:Euler}), we arrive at
\[
\frac{{\delta} L}{{\delta}\Phi}=i\omega\tau^{\prime}\bar{\Phi}+i\omega\tau\bar{\Phi}^{\prime},
\]
which leads to the following flow equation:
\begin{equation}
\tau^{\prime}=\left\llbracket \frac{{\delta} L}{{\delta}\Phi}\right\rrbracket \left\llbracket i\omega\bar{\Phi}\right\rrbracket^{-1}-\tau\left\llbracket \bar{\Phi}^{\prime}\right\rrbracket \left\llbracket \bar{\Phi}\right\rrbracket^{-1}.\label{eq:raw}
\end{equation}
Note that from (\ref{eq:Hamilton}) we have
\begin{equation}\label{canonical}
\tau=\left\llbracket \frac{{\delta} L}{{\delta}\Phi^{\prime}}\right\rrbracket \left\llbracket i\omega\bar{\Phi}\right\rrbracket ^{-1},
\end{equation}
so generically both terms on the right hand side of (\ref{eq:raw})
can be expressed in terms of $\tau$ and some {matrices independent of perturbations},
which gives rise to the following form
\begin{equation}\label{main}
\tau^{\prime}=M-N\tau-\tau\tilde{N}-\tau O\tau
\end{equation}
of the flow equation. Here $M$, $N$, $\tilde{N}$ and $O$ are  {matrices independent of perturbations}. In this way, the flow equation of $\tau$ can be deduced
without explicitly using the equation of motion, although (\ref{eq:Euler})
does be implicitly used.

An explicit derivation of the flow equation for the RN-AdS bulk spacetime in the Einstein-Maxwell theory using the above formalism is provided in the Appendix, where one can see that Onsager's reciprocal relation is reflected in this formalism. In practice, taking the canonical definition (\ref{canonical}) of $\tau$ is not necessary to calculate the transport coefficients that we are interested in, which can be seen from the following discussion of the holographic model with momentum dissipation. In this case, the matrix $\tau$ {(denoted as $\tilde{\tau}$ there)} has no definite symmetry and Onsager's reciprocal relation is not explicitly reflected. {Meanwhile, without considering the effective bulk perturbation action there, we can directly deduce the flow equation of $\tilde{\tau}$ using the equations of motion for the linear perturbations.}

{Our flow equation (\ref{main}) of $\tau$ can be regarded as a generalization of the flow equation in \cite{liu1} to the case of mixed (coupled) transportation. As we will see in the following sections, our matrix flow equation provides a novel approach to compute the transport coefficients of the boundary field theory in both the DC and AC cases, independent of the precise connection between the $r$ flow of transport coefficients and the real RG flow of the boundary field theory \cite{liu1}.}

\section{{Holographic transport with momentum dissipation}}
We consider a general class of Einstein-Maxwell-Dilaton theories with linear axion fields
\bea\label{uniaction}
S=\int \! d^{4}x \!\sqrt{-g} \bigg[\bigg(R-V(\phi)-\frac{1}{2}\partial \phi^2-\frac{1}{2}Y(\phi)\!\sum^{2}_{i=1}\partial\chi^2_{i}\bigg)-\frac{1}{4}Z(\phi)F^2\bigg].
\eea
where $R$ is the Ricci scalar and $\chi_i$ is a collection of $2-$massless linear axions.
The action consists of the Einstein gravity, axion fields, a Maxwell field and a dilaton field.

The action yields equations of motion as
\bea
0&=&R_{\mu\nu}-\frac{1}{2}\partial_{\mu}\phi \partial_{\nu} \phi-\frac{Y(\phi)}{2}\sum^{2}_{i}\partial_{\mu}\chi_i \partial_{\nu} \chi_i-\frac{V(\phi)}{2}g_{\mu\nu}-\frac{Z(\phi)}{2}F_{\mu}^{\rho}F_{\nu\rho}+\frac{Z(\phi)}{8}F^2g_{\mu\nu},\\
0&=&\nabla_{\mu}\bigg(Z(\phi)F^{\mu\nu}\bigg),\\
0&=&\nabla_{\mu}\bigg(Y(\phi)\nabla^{\nu}\chi_i\bigg),\\
0&=&\Box\phi-V'(\phi)-\frac{1}{4}Z'(\phi)F^2-\frac{1}{2}Y'(\phi)\sum^{2}_{i}(\partial \chi_i)^2.
\eea

This model has been widely studied recently and several different types of black hole solution has been obtained. In what follows, we mainly focus on
the simple black hole solution to the action (\ref{uniaction}) taking the form \cite{withers}
\bea
ds^2&=&-f(r)dt^2+\frac{dr^2}{f(r)}+r^2 dx^2+r^2 dy^2,\\
f(r)&=&r^2-\frac{\beta^2}{2}-\frac{m}{r}+\frac{\mu^2 \rh^2}{4 r^2},\qquad m=\rh^3 \left(1+\frac{\mu^2}{4\rh^2}-\frac{\beta^2}{2\rh^2}\right),\\
A_t&=&\mu \left(1-\frac{\rh}{r}\right)dt,\qquad\chi_i=\beta x_i,
\eea
where $\mu$ is the chemical potential and $\rh$ is the event horizon radius. This is an asymptotic AdS solution as $r\rightarrow \infty$. The temperature of the dual field theory is given by
\be
T=\frac{1}{4\pi}\left(3\rh-\frac{\mu^2+2\beta^2}{4\rh}\right).
\ee
The entropy density reads $s=4 \pi\rh^2$. The black hole solution was first proposed and studied in \cite{withers}. The AC conductivity with/without external magnetic field was carried out in \cite{kim,kim2}. In this paper, we revisit this model and calculate the transport coefficients by using the formalism proposed in section 2 as an explicit demonstration of the powerfulness of our method.

\subsection{DC transport coefficients}
For the purpose of calculating the response of this system, we consider linear perturbation as follows
\begin{eqnarray} \label{flucA}
\delta A_x(t,r) &= \int^{\infty}_{-\infty} \frac{d \omega}{2\pi}  e^{-i\omega t}  a_{x}(\omega,r)\,, \\
\delta g_{tx}(t,r) &= \int^{\infty}_{-\infty} \frac{d \omega}{2\pi} e^{-i\omega t} r^2 h_{tx}(\omega,r)\,,  \label{flucg} \\
\delta \chi_x(t,r) &= \int^{\infty}_{-\infty} \frac{d \omega}{2\pi} e^{-i\omega t} \chi_{x}(\omega,r)/\beta \,.\label{flucPsi}
\end{eqnarray}
The equations of motion for these linear perturbations take the form
\bea
0&=&a''_x+\frac{f'}{f}a'_x+\frac{\omega^2}{f^2}a_x+\frac{\mu \rh}{f}h'_{tx},\label{eqax}\\
0&=& \chi''+\left(\frac{f'}{f}+\frac{2}{r}\right)\chi'+\frac{\omega^2}{f^2}\chi-\frac{i\omega \beta^2}{f^2}h_{tx},\label{eqhtx}\\
0&=& h''_{tx}+\frac{4}{r}h'_{tx}+\frac{\mu \rh a'_x}{r^4}-\frac{i\rh^2 \omega \chi}{r^2 f}-\frac{\beta^2 h_{tx}}{r^2 f},\label{eqchi}\\
0&=& \frac{i \omega r^2}{f}h'_{tx}+\frac{i\omega \mu \rh}{r^2 f}a_x-\chi'.\label{connection}
\eea
%From the regularity condition at the event horizon, we know that $a''_x$,  $h''_{tx}$ and $\chi''$ are finite at the event horizon, but $a'_x$ and $h_{tx}$ are not regular at the event horizon.
Imposing regularity condition on (\ref{eqchi}), we obtain
\be\label{reguhtx}
h_{tx}|_{r=\rh}=\bigg(\frac{\mu f a'_x}{\beta^2 \rh}-i\omega \rh^2 \chi\bigg)\bigg|_{r=\rh}.
\ee
By eliminating $h_{tx}$ and introducing $\phi=r^2f\chi'/(i\omega)$, we are able to rewrite (\ref{eqax}) and (\ref{eqhtx})  as
\bea
(fa_{x}^{\prime})^{\prime} & = & Aa_{x}+B\phi,\label{one}\\
(r^{-2}f\phi^{\prime})^{\prime} & = & Ca_{x}+D\phi,\label{two}
\eea
where
\[
A=\mu^{2}\frac{\rh^{2}}{r^{4}}-\frac{\omega^{2}}{f},\qquad B=-\mu\frac{\rh}{r^{4}},
\]
\[
C=-\beta^{2}\mu\frac{\rh}{r^{4}},\qquad D=\frac{\beta^{2}}{r^{4}}-\frac{\omega^{2}}{r^2 f}.
\]
Multiplying (\ref{two}) with ${\mu \rh}/{\beta^2}$ and combining with (\ref{one}), we obtain
\be
-\left(fa'_x+\frac{\rh \mu}{r^2\beta^2}f\phi'\right)'=\frac{\omega^2}{f}a_x+\frac{\omega^2 \rh \mu}{\beta^2 r^2 f}\phi.
\ee
Therefore, to order $\mathcal{O}(\omega)$, we can define a radially conserved electrical current
\be \label{jx1}
J_x=-fa'_x-\frac{\rh \mu}{r^2\beta^2}f\phi'.
\ee
{Up to order $\mathcal{O}(\omega)$, we can see from (\ref{eqhtx})} that (\ref{jx1}) is equal to the radially conserved electrical current from the Maxwell equation
\be\label{jx2}
J_x=-fa'_x-\mu \rh h_{tx}.
\ee
We then try to {use the method described in the previous section} to calculate the transport coefficients. {However, instead of using the canonical definition (\ref{canonical}) of the transport matrix $\tau$, we define} a
matrix $\tilde{\tau}$ from
\[
\left \llbracket \begin{array}{c}
-fa_{x}^{\prime}\\
-r^{-2}f\phi^{\prime}
\end{array}\right \rrbracket=\tilde{\tau}\left \llbracket\begin{array}{c}
i\omega a_{x}\\
i\omega\phi
\end{array}\right \rrbracket. \]
That is to say, $\tilde{\tau}$ is defined as
\begin{eqnarray}
\tilde{\tau}\equiv\left \llbracket \begin{array}{c}
-r^{d-3}fa_{x}^{\prime}\\
-r^{1-d}f\phi^{\prime}
\end{array}\right \rrbracket\left \llbracket\begin{array}{c}
i\omega a_{x}\\
i\omega\phi
\end{array}\right \rrbracket^{-1}.
\end{eqnarray}
{Recall that t}he special notation $\llbracket ... \rrbracket$ denotes that it is a matrix, not a vector.
The radial derivative of the matrix $\tilde{\tau}$ can be expressed as
\begin{eqnarray}
\tilde{\tau}^{\prime} & = & \left \llbracket \begin{array}{c}
-fa_{x}^{\prime}\\
-r^{-2}f\phi^{\prime}
\end{array}\right \rrbracket^{\prime}\left \llbracket\begin{array}{c}
i\omega a_{x}\\
i\omega\phi
\end{array}\right \rrbracket^{-1}-i\omega\tilde{\tau}\left \llbracket\begin{array}{c}
a_{x}\\
\phi
\end{array}\right \rrbracket^{\prime}\left \llbracket\begin{array}{c}
i\omega a_{x}\\
i\omega\phi
\end{array}\right \rrbracket^{-1}\nonumber\\
 & = & \left \llbracket\begin{array}{c}
-(Aa_{x}+B\phi)\\
-(Ca_{x}+D\phi)
\end{array}\right \rrbracket  \left \llbracket\begin{array}{c}
i\omega a_{x}\\
i\omega\phi
\end{array}\right \rrbracket^{-1}-i\omega\tilde{\tau}\left \llbracket\begin{array}{c}
a_{x}^{\prime}\\
\phi^{\prime}
\end{array}\right \rrbracket\left \llbracket\begin{array}{c}
i\omega a_{x}\\
i\omega\phi
\end{array}\right \rrbracket^{-1}\nonumber\\
 & = & -\frac{1}{i\omega}\left(\begin{array}{cc}
A & B\\
C & D
\end{array}\right)+i\omega\tilde{\tau}\left(\begin{array}{cc}
f^{-1} & 0\\
0 & (r^{-2}f)^{-1}
\end{array}\right)\tilde{\tau}.\label{floweq}
\end{eqnarray}
This is the main equation. Remarkably, we express the flow equation in terms of matrices and this is very different from the RG flow method investigated in \cite{sken,ia1,ia2}, where only uncoupled flow equations are considered.

Multiplying equation (\ref{floweq}) with $f$, we obtain
\[
f\tilde{\tau}^{\prime}=-\frac{f}{i\omega}\left(\begin{array}{cc}
A & B\\
C & D
\end{array}\right)+i\omega\tilde{\tau}\left(\begin{array}{cc}
1 & 0\\
0 & r^{2}
\end{array}\right)\tilde{\tau}.
\]
At the event horizon, we have
\[
0=\left(\begin{array}{cc}
-i\omega & 0\\
0 & -i \rh^{-2}\omega
\end{array}\right)+i\omega\tilde{\tau}_\mathrm{H}\left(\begin{array}{cc}
1 & 0\\
0 & \rh^{2}
\end{array}\right)\tilde{\tau}_\mathrm{H}.
\]
Regularity at the event horizon yields
\[
\tilde{\tau}_\mathrm{H}=\left(\begin{array}{cc}
1 & 0\\
0 & \rh^{-2}
\end{array}\right).
\]
Substituting the expression of $\tilde{\tau}_{H}$ into the definition of $\tilde{\tau}$, one immediately obtain
\begin{eqnarray}
-fa_{x}^{\prime} & \to & i\omega a_{x} \big|_{r=\rh},\label{eq:a_x}\\
-f\phi^{\prime} & \to & i\omega\phi \big|_{r=\rh},\label{eq:phi}
\end{eqnarray}
near the horizon. Together with the event horizon regularity condition from (\ref{one}) or ($\ref{two}$) in the zero frequency limit
\be
\phi=\mu \rh a_x.
\ee
The conserved electrical current (\ref{jx1}) can be recast as
\bea
J_x &=&-fa'_x-\frac{\rh \mu}{r^2\beta^2}f\phi'\nonumber\\
&=&i\omega a_x+i \omega \frac{\mu^2}{\beta^2}a_x.
\eea
Therefore, the DC conductivity can be evaluated at the event horizon
\be\label{dc}
\sigma_{DC}=\frac{\partial J_x}{\partial E_x}=1+\frac{\mu^2}{\beta^2},
\ee
where $E_x=i\omega a_x$. \footnote{{An alternative} derivation of the DC electric conductivity can be achieved by using (\ref{jx2}) and the in-going boundary condition
$a_x=(r-\rh)^{-i \omega/4\pi T}a^0_x$ with $a^0_x$ a constant.
Together with the regularity condition from (\ref{eqchi}) at zero frequency:
$h_{tx}=\frac{\mu f a'_x}{\beta^2\rh}$, we can also obtain the DC conductivity. In this sense, {here the use of $\tilde{\tau}$ is not necessary, but only for illustration of determining the horizon boundary condition $\tilde{\tau}_\mathrm{H}$ from the flow equation (\ref{floweq})}. But as we will see later, the $\tilde{\tau}$ matrix presented in (\ref{floweq}) can greatly simplify the computation of the AC conductivity{, where no analytical methods are available}.}

\subsection{Thermoelectric and thermal conductivities}
In this subsection, we consider electric conductivity at zero heat current as first input condition and then evaluate thermoelectric and thermal conductivities via the general response theory.

Considering $(\ref{eqchi})$ together with equations of motion of the Maxwell fields to the linear order, we can write down a radially conserved heat current
\be \label{heatcurrent}
\mathcal{Q}_x=\sqrt{\frac{g_{tt}}{g_{rr}}}(-g^{tt}h_{tx}\partial_r g_{tt}+h'_{tx})-A_{t} J_x.
\ee
After imposing the regularity condition at the event horizon, we are able to evaluate the heat current
\be
\mathcal{Q}_x=i \frac{4\pi T \mu \rh}{\beta^2} \omega a_x,
\ee
where we have used the fact $A_t(\rh)=0$ at the event horizon.
The thermoelectric conductivity  can then be expressed as
\be
\alpha=\frac{\partial \mathcal{Q}_x}{T \partial E_x}=\frac{4\pi  \mu \rh}{\beta^2}.
\ee
The thermal conductivity satisfies the following ansatz
\bea
J_x&=& \sigma E_{x}-\alpha \nabla_x T,\\
\mathcal{Q}_x &=& \bar{\alpha}T E_x-\bar{\kappa}\nabla_x T.
\eea
Note that zero heat current $\mathcal{Q}_x =0$ gives us
\be
\nabla_x T=\frac{\bar{\alpha} T}{\bar{\kappa}}E_x.
\ee
The electric current with vanishing heat current now is mainly contributed by the pair-production of the boundary vacuum
\be
J_x=\sigma E_{x}-\frac{\alpha \bar{\alpha} T}{\bar{\kappa}}E_x.
\ee
Therefore, we have
\be
\frac{\partial J_x}{\partial E_x}=\sigma-\frac{\alpha \bar{\alpha} T}{\bar{\kappa}}=\sigma^0,
\ee
where $\sigma^0=1$. The universal formula of electrical conductivity with vanishing heat current for isotropic and anisotropic systems has been formulated in \cite{gsw2015}
\be
\prod_i \sigma^{0}_{ii}|_{q_i=0}=\mathcal{A}^{d-3} \prod_i Z_i^{d-1}(\rh),
\ee
where $\mathcal{A}$ and $Z_i$ are area density per unit volume of the black hole horizon and gauge field couplings, respectively. Here we consider $\sigma^{0}_{ii}$ as an already known physical quantity.
The thermal conductivity then reads
\be
\bar{\kappa}=\frac{\alpha \bar{\alpha} T}{\sigma-\sigma^0}=\frac{16 \pi^2 T \rh^2}{\beta^2}.
\ee
We can see that once the conserved currents are obtained, the linear responses are easily calculated.

\subsection{DC conductivity flow and Optical conductivity}\label{AC}
\begin{figure}
\begin{center}
	\includegraphics[scale=0.5]{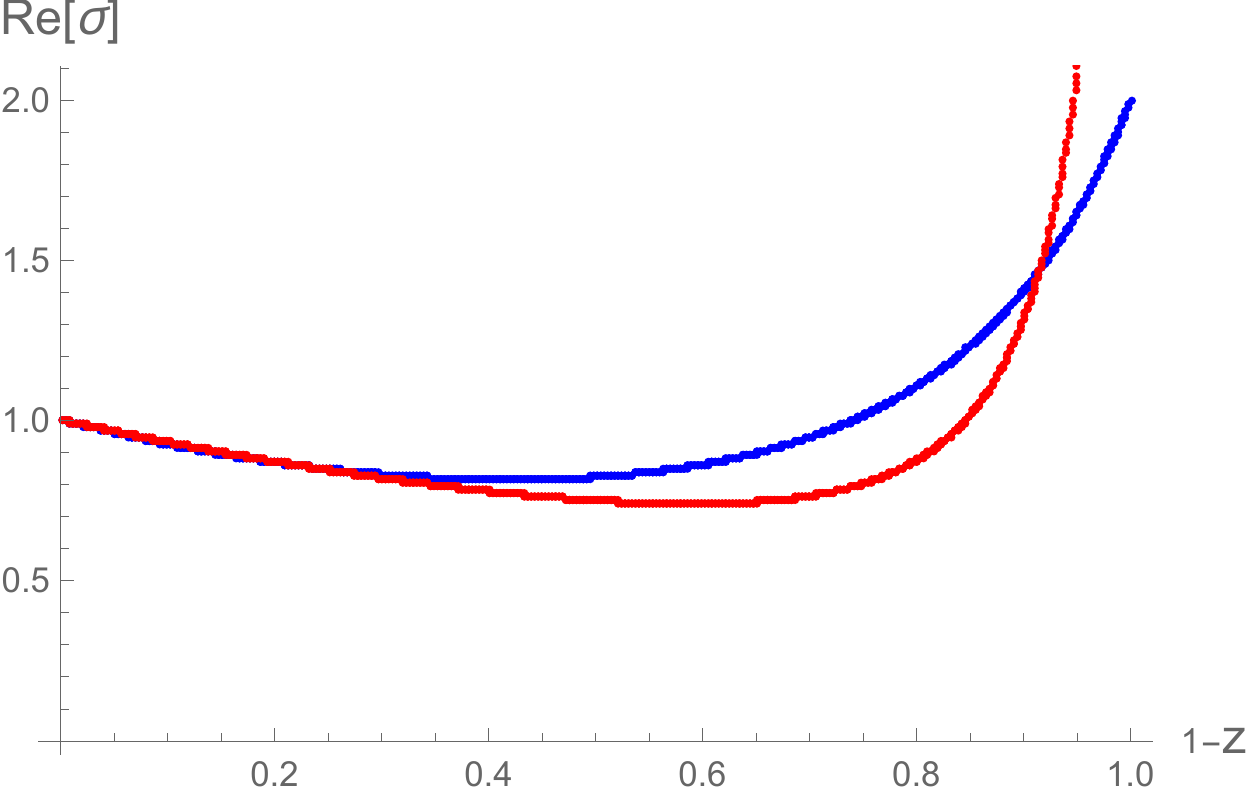}
	\includegraphics[scale=0.5]{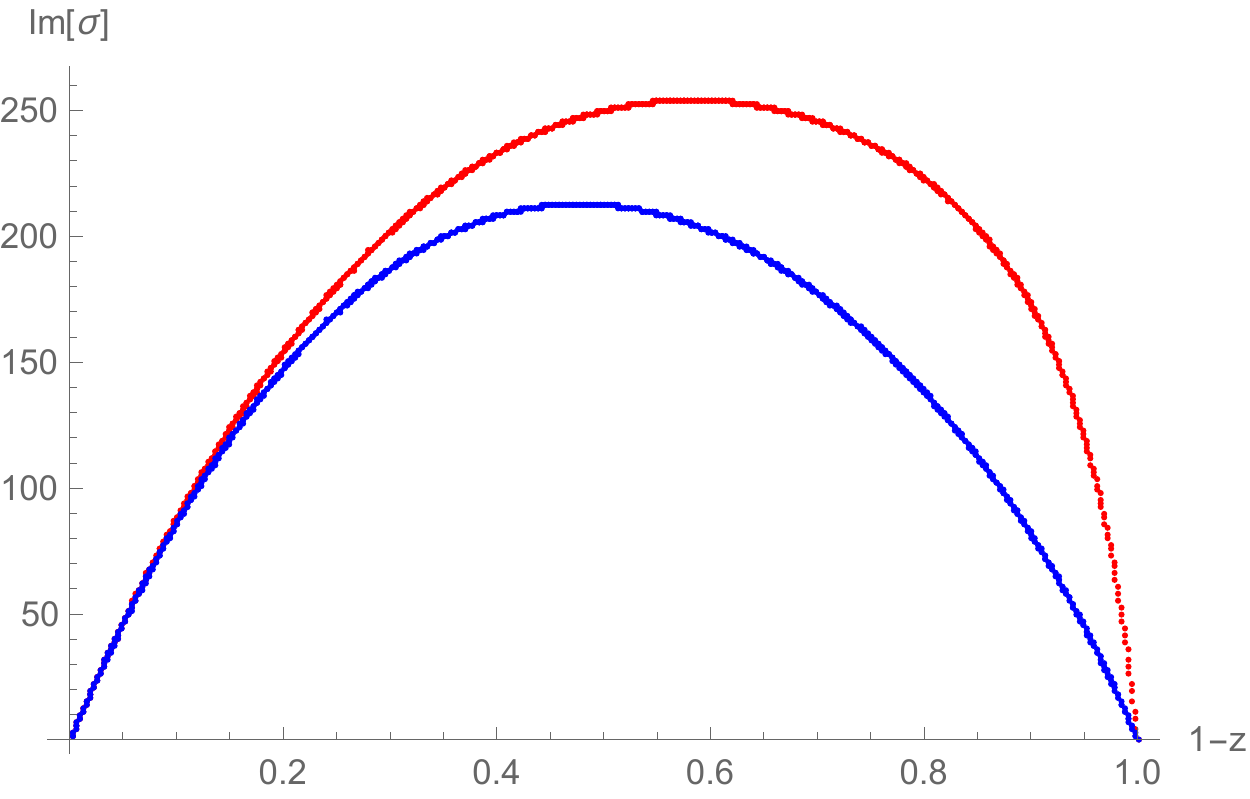}
\end{center}
\caption{The radial flow of the optical conductivity for $\beta/\mu=1/2$ ({red} line) and $\beta/\mu=1$ ({blue} line). Here we set $\rh=1$ and $\omega=0.001$. {On the boundary $z\to 0$, the blue line tends to $\mathrm{Re}(\sigma)\approx 2$ and the red line (not fully shown in the figure) tends to $\mathrm{Re}(\sigma)\approx 5$, which agrees with the DC conductivity (\ref{dc}).}}\label{fflow}
\end{figure}
With the main flow equation (\ref{floweq}) at hand, we are able to evaluate the conductivity at a finite cutoff surface numerically. For the real part of the conductivity, we can see from Figure \ref{fflow} that there is a non-trivial flow from the event horizon to the boundary. At the event horizon $\sigma(\rh)=1$, while near the boundary $r\rightarrow \infty$ ($z\rightarrow 0$) the DC conductivity is reduced to (\ref{dc}){, which can be viewed as a valuable check of the analytical calculation of the DC conductivity, as well as a check of our matrix flow formalism (\ref{floweq}) itself}. In particular, it is noteworthy that the imaginary part $\mathrm{Im}(\sigma_{DC})$ of the DC conductivity goes through a non-trivial flow in the radial direction, but always drops to zero at the conformal boundary.

\begin{figure}
\begin{center}
	\includegraphics[scale=0.5]{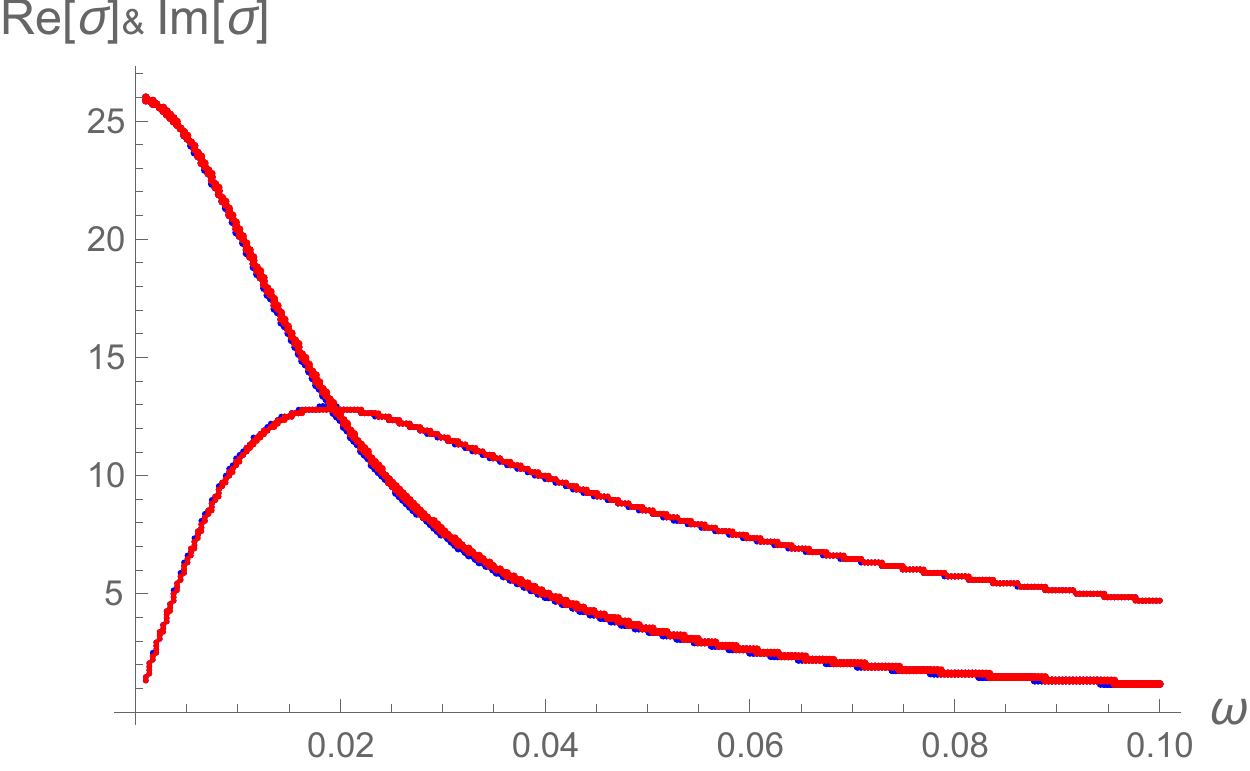}
	\includegraphics[scale=0.5]{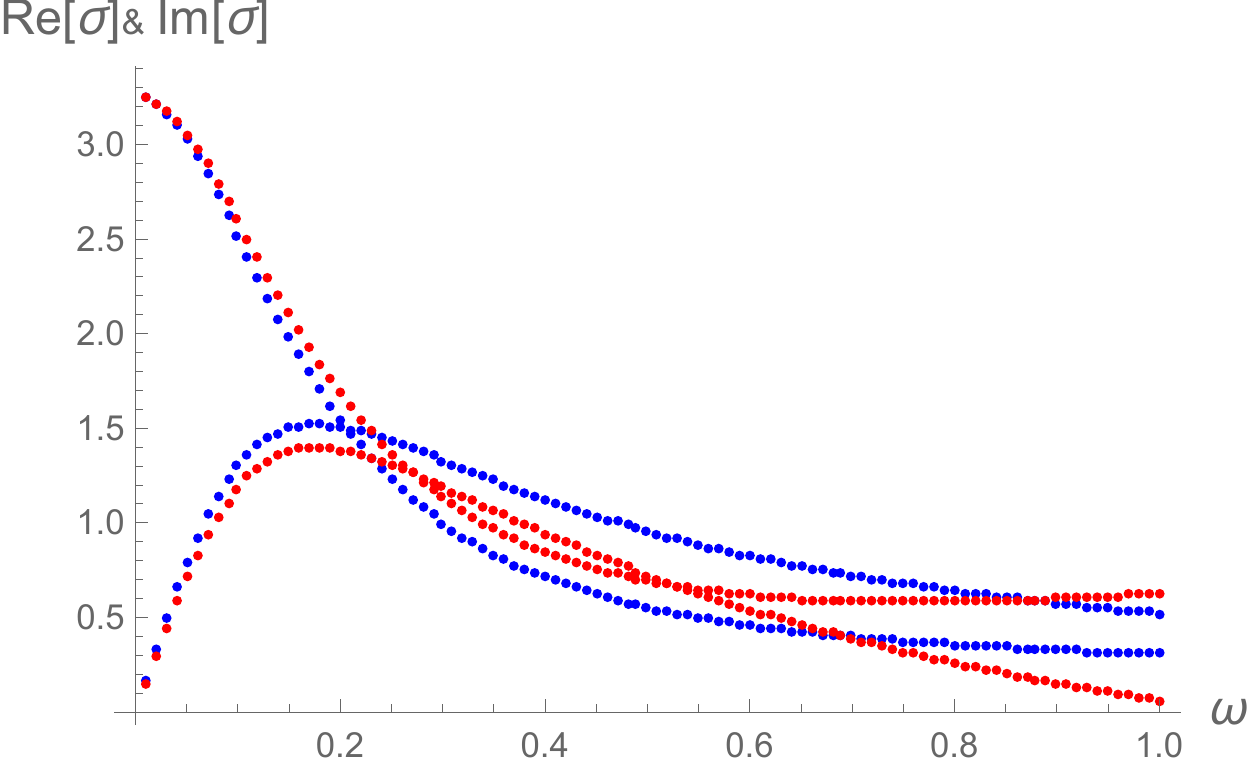}
	\includegraphics[scale=0.5]{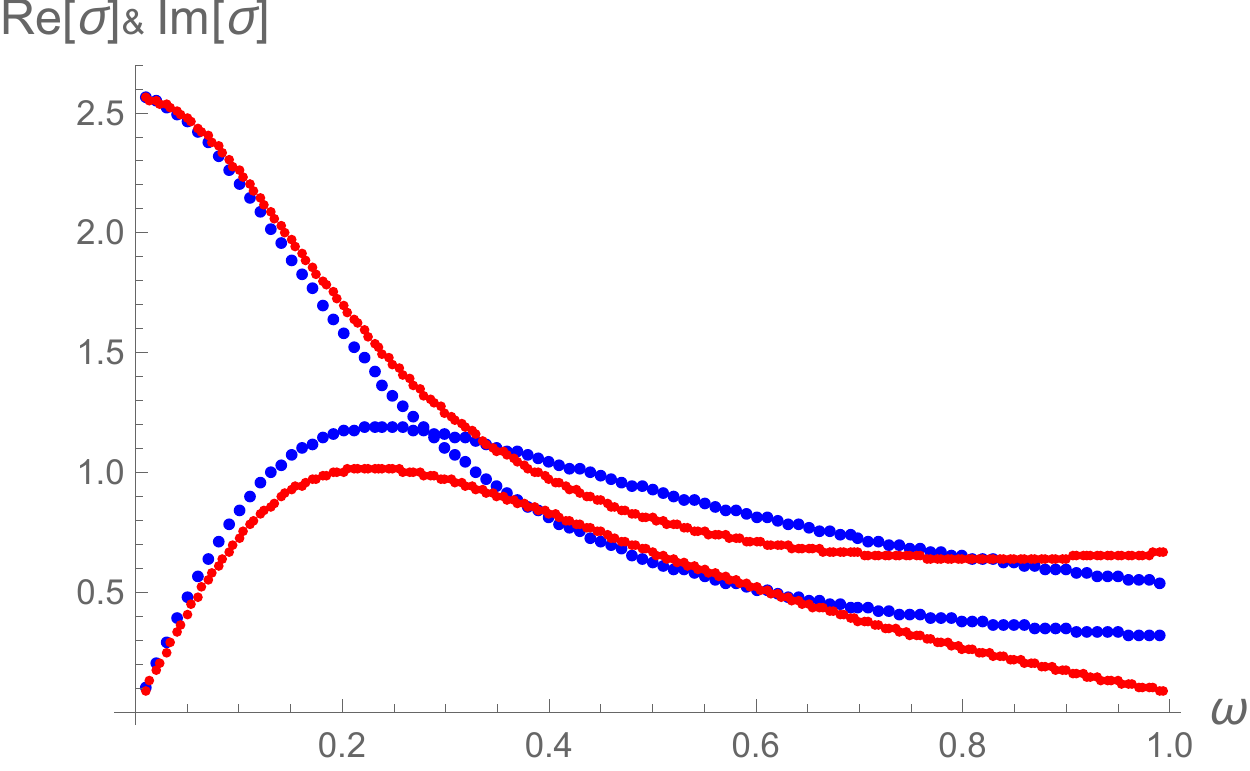}
	\includegraphics[scale=0.5]{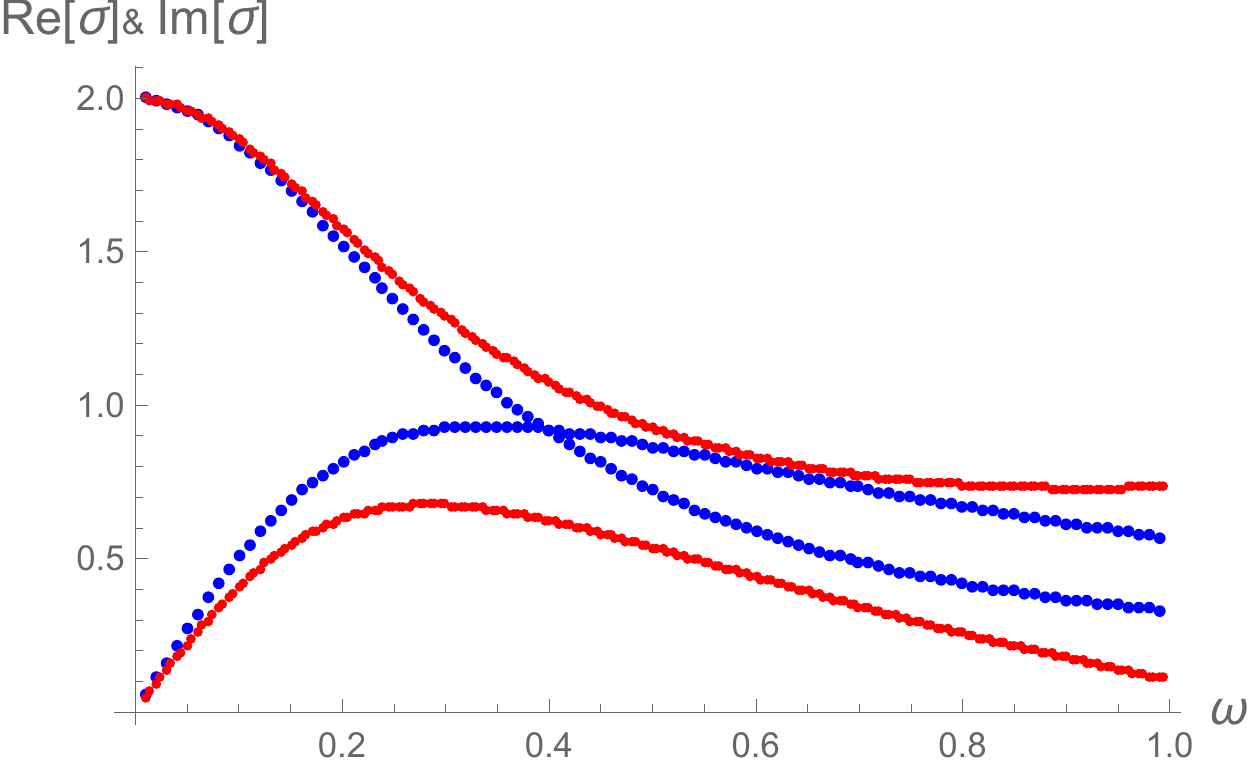}
\end{center}
\caption{The optical conductivity as a function of frequency for $\beta/\mu=1/5,2/3,4/5,1$ from left to right. The red dots are data, and the blue lines are fit to the Drude formula.\label{fig1}}
\end{figure}

Since the flow equation (\ref{floweq}) is simply a first-order non-linear ordinary differential equation, one can compute the optical conductivity more easily.
As shown in Figure \ref{fig1}, for small $\beta/\mu$, the numerical computation shows that the optical conductivity takes the form of the Drude conductivity as
\be
\sigma(\omega)=\sigma_{Q}+\frac{K \varsigma}{1-i\omega \varsigma},
\ee
where
\begin{equation}\label{}
\sigma_{Q}=(\frac{Ts}{Ts+\mu q})^2,\qquad K=\frac{q^2}{Ts+\mu q},\qquad \varsigma=\frac{\sigma_{DC}-\sigma_{Q}}{K}
\end{equation}
with $\sigma_{DC}$ given by (\ref{dc}).

Since the translational symmetry is broken, the $1/\omega$ pole in the imaginary part disappears. From Figure \ref{fig1}, we can see that as $\beta/\mu$ becomes bigger, the maximum value of the peak in the real part goes down. The numerical results are in good agreement with the DC conductivity given in (\ref{dc}). We also note that as $\beta/\mu$ increases deviations from the Drude model can be observed and the conductivity become incoherent without a Drude peak. The numerical results are totally in agreement with the numerical studies obtained in \cite{kim}. The advantages of the RG flow method shown here are that it does not rely on the {inconvenient} shooting method or the {resource-consuming} pseudo-spectral method, because the {flow equation of $\bar{\tau}$} is simply a first order ordinary differential equation {with a single boundary condition at the horizon,} for which a simple and completely controllable Runge-Kutta marching from the horizon to the boundary is sufficient.

\subsection{DC transport at finite magnetic field}
In this section, we will show explicitly how to calculate the conductivity analytically in the presence of an external magnetic field by using the RG flow approach.
After turning on the magnetic field, the metric is given by
\begin{eqnarray}\label{dyon02}
&&ds^2 = -f(r) dt^2 +\frac{1}{f(r)} d r^2 + r^2 \left(d x^2 +  dy^2\right) \,, \\
&& f(r)= r^2 - \frac{ \beta^2}{ 2 } - \frac{m}{r} + \frac{ \mu^2 + q_m^2 }{4} \frac{\rh^{2}}{r^{2}} \,, \\
&&a(r) = \mu \left( 1- \frac{\rh}{r} \right) \,, \qquad B = q_m \rh \,, \\
&&\chi_1= \beta x\,, \qquad \chi_2 = \beta y \,,
\end{eqnarray}
where $\rh$ is the location of the the brane horizon and
\begin{eqnarray}
m = \rh^3 \left( 1+\frac{\mu^2 + q_m^2}{4 \rh^2} - \frac{\beta^2}{2 \rh^2} \right) \,.
\end{eqnarray}
The Hawking temperature is given by
\be
T=\frac{1}{4\pi}\bigg(3\rh-\frac{\mu^2+q^2_m+2\beta^2}{4\rh}\bigg),
\ee
The linear perturbations $\delta g_{ti}$, $\delta A_i$ and $\delta \chi_i$ can be expressed in momentum space as
\begin{eqnarray}
\delta A_i(t,r) &=& \int^{\infty}_{-\infty} \frac{d \omega}{2\pi} e^{-i\omega t} a_{i}(\omega,r)\,, \\
\delta g_{ti}(t,r) &=& \int^{\infty}_{-\infty} \frac{d \omega}{2\pi} e^{-i\omega t} r^2 h_{ti}(\omega,r)\,, \label{flucg1} \\
\delta \chi_i(t,r) &=& \int^{\infty}_{-\infty} \frac{d \omega}{2\pi} e^{-i\omega t} \psi_{i}(\omega,r) \,,\label{flucPsi}
\end{eqnarray}
where $r^2$ in the metric fluctuation (\ref{flucg1}) is introduced to make an asymptotic solution of $h_{ti}$ constant at the boundary $r \rightarrow \infty$.
The linearised equations for the Fourier components are given as follows.

The linearized Einstein equations are given by
\begin{eqnarray} \label{ee1}
\frac{q_m^2 h_{ti}}{r^4 f}+\epsilon_{ij}\frac{i \omega q_m  a_j}{r^4 f}+\frac{\beta^2 h_{ti}}{r^2 f}+\frac{i \beta \omega \psi_i}{r^2 f}-\frac{\mu a_i'}{r^4}-\frac{4  h_{ti}'}{r}-h_{ti}''&=&0 \,, \\
\epsilon_{ij}\frac{i f q_m a_j'}{r^4 \omega }+\frac{i \beta f \psi_i'}{r^2 \omega }+\epsilon_{ij}\frac{i \mu q_m h_{{tj}}}{r^4 \omega }+\frac{\mu a_i}{r^4}+h_{ti}'&=&0 \,,
\end{eqnarray}
where $\epsilon_{ij}$ is the Levi-Civita symbol.
The linearized Maxwell equation yields the form
\begin{eqnarray} \label{me1}
\frac{f' a_i'}{f}+\epsilon_{ij}\frac{i \omega q_m h_{{tj}}}{f^2}+\frac{\mu h_{ti}'}{f}+\frac{\omega^2 a_i}{f^2}+ a_i''&=&0 \,.
\end{eqnarray}
Similarly we can write down the linearized axion equation:
\begin{eqnarray} \label{se1}
\frac{f' \psi_i'}{f}-\frac{i \beta \omega h_{ti}}{f^2}+\frac{\omega^2 \psi_i}{f^2}+\frac{2 \psi_i'}{r}+\psi_i''&=&0 \,.
\end{eqnarray}
It is difficult to solve equations (\ref{ee1})-(\ref{se1}) directly, although among these eight equations, only six are independent. We notice that even in the presence of magnetic field,
the spatial $SO(2)$ symmetry is unbroken and it is convenient to organize the fields as
\begin{eqnarray}
a_z=(a_x-i a_y)/2,\qquad h_{tz}=(h_{tx}-i h_{ty})/2,\qquad \psi_z=(\psi_x-i \psi_y)/2.
\end{eqnarray}
The equations of motion can be rewritten as
\begin{equation}
(fa_{z}^{\prime})^{\prime}+\mu h_{tz}^{\prime}+\frac{\omega^{2}a_{z}}{f}-\frac{\omega q_{m}h_{tz}}{f}=0,\label{eq:Maxwell}
\end{equation}
\begin{equation}
(r^{2}f\psi_{z}^{\prime})^{\prime}-\frac{i\beta\omega r^{2}h_{tz}}{f}+\frac{\omega^{2}r^{2}\psi_{z}}{f}=0,\label{eq:Klein-Gordon}
\end{equation}
\begin{equation}
-q_{m}fa_{z}^{\prime}+i\beta r^{2}f\psi_{z}^{\prime}-\mu q_{m}h_{tz}+\omega\mu a_{z}+\omega r^{4}h_{tz}^{\prime}=0,\label{eq:Einstein}
\end{equation}
\begin{equation}
\frac{q_{m}^{2}h_{tz}}{f}-\frac{\omega q_{m}a_{z}}{f}+\frac{\beta^{2}r^{2}h_{tz}}{f}+\frac{i\beta\omega r^{2}\psi_{z}}{f}-\mu a_{z}^{\prime}-(r^{4}h_{tz}^{\prime})^{\prime}=0. \label{htx}
\end{equation}
{Using the standard procedure by imposing boundary conditions at the boundary and the horizon, we can solve equations (\ref{eq:Maxwell}-\ref{eq:Einstein}) as it was done in \cite{kim2}. However, for the purpose of consistency and showing the power of the $\tilde{\tau}$ matrix, we follow the analysis presented in section (3.1) and derive first-order ordinary differential equations of motion in the presence of magnetic field. \footnote{Using the same flow equation, one can also easily compute the AC transport coefficients in the presence of magnetic field as in section \ref{AC}, if needed.}}
One can see that from (\ref{eq:Maxwell}), (\ref{eq:Klein-Gordon}) and (\ref{htx}), one can naturally achieve (\ref{eq:Einstein}).
So let us define
\begin{eqnarray}\label{definesigma}
\left\llbracket\begin{array}{c}
-fa_{z}^{\prime}\\
-r^{2}f\psi_{z}^{\prime}\\
-r^{4}h_{tz}^{\prime}
\end{array}\right\rrbracket=\tilde{\tau}\left\llbracket\begin{array}{c}
i\omega a_{z}\\
i\omega\psi_{z}\\
i\omega h_{tz}
\end{array}\right\rrbracket.
\end{eqnarray}
The flow equation of $\tilde{\tau}$ can be expressed as
\begin{eqnarray}
\tilde{\tau}^\prime & = & \left\llbracket\begin{array}{c}
-fa_{z}^{\prime}\\
-r^{2}f\psi_{z}^{\prime}\\
-\frac{q_{m}}{\omega}fa_{z}^{\prime}+\frac{i\beta}{\omega}r^{2}f\psi_{z}^{\prime}-\frac{\mu q_{m}}{\omega}h_{tz}+\mu a_{z}
\end{array}\right\rrbracket^{\prime}\left\llbracket\begin{array}{c}
i\omega a_{z}\\
i\omega\psi_{z}\\
i\omega h_{tz}
\end{array}\right\rrbracket^{-1}-\tilde{\tau}\left\llbracket\begin{array}{c}
i\omega a_{z}\\
i\omega\psi_{z}\\
i\omega h_{tz}
\end{array}\right\rrbracket^{\prime}\left\llbracket\begin{array}{c}
i\omega a_{z}\\
i\omega\psi_{z}\\
i\omega h_{tz}
\end{array}\right\rrbracket^{-1}\nonumber\\
% & = & \left(\begin{array}{c}
%\mu h_{tz}^{\prime}+\frac{\omega^{2}a_{z}}{f}-\frac{\omega q_{m}h_{tz}}{f}\\
%-\frac{i\beta\omega r^{2}h_{tz}}{f}+\frac{\omega^{2}r^{2}\psi_{z}}{f}\\
%\frac{q_{m}}{\omega}(\mu h_{tz}^{\prime}+\frac{\omega^{2}a_{z}}{f}-\frac{\omega q_{m}h_{tz}}{f})+\frac{i\beta}{\omega}(\frac{i\beta\omega r^{2}h_{tz}}{f}-\frac{\omega^{2}r^{2}\psi_{z}}{f})-\frac{\mu q_{m}}{\omega}h_{tz}^{\prime}+\mu a_{z}^{\prime}
%\end{array}\right)\left(\begin{array}{c}
%i\omega a_{z}\\
%i\omega\psi_{z}\\
%i\omega h_{tz}
%\end{array}\right)^{-1}\\
% &  & -\tilde{\sigma}\left(\begin{array}{c}
% i\omega a_{z}^{\prime}\\
% i\omega\psi_{z}^{\prime}\\
% i\omega h_{tz}^{\prime}
% \end{array}\right)\left(\begin{array}{c}
% i\omega a_{z}\\
% i\omega\psi_{z}\\
% i\omega h_{tz}
%\end{array}\right)^{-1}\\
 & = & \left\llbracket\begin{array}{c}
\mu h_{tz}^{\prime}+\frac{\omega^{2}a_{z}}{f}-\frac{\omega q_{m}h_{tz}}{f}\\
-\frac{i\beta\omega r^{2}h_{tz}}{f}+\frac{\omega^{2}r^{2}\psi_{z}}{f}\\
\frac{q_{m}\omega a_{z}}{f}-\frac{q_{m}^{2}h_{tz}}{f}-\frac{\beta^{2}r^{2}h_{tz}}{f}-\frac{i\beta\omega r^{2}\psi_{z}}{f}+\mu a_{z}^{\prime}
\end{array}\right\rrbracket\left\llbracket\begin{array}{c}
i\omega a_{z}\\
i\omega\psi_{z}\\
i\omega h_{tz}
\end{array}\right\rrbracket^{-1}-\tilde{\tau}\left(\begin{array}{ccc}
-\frac{i\omega}{f} & 0 & 0\\
0 & -\frac{i\omega}{r^{2}f} & 0\\
0 & 0 & -\frac{i\omega}{r^{4}}
\end{array}\right)\tilde{\tau}\nonumber\\
 & = & \left\llbracket\begin{array}{c}
\frac{\omega^{2}a_{z}}{f}-\frac{\omega q_{m}h_{tz}}{f}\\
-\frac{i\beta\omega r^{2}h_{tz}}{f}+\frac{\omega^{2}r^{2}\psi_{z}}{f}\\
\frac{q_{m}\omega}{f}a_{z}-\frac{q_{m}^{2}+\beta^{2}r^{2}}{f}h_{tz}-\frac{i\beta\omega r^{2}}{f}\psi_{z}
\end{array}\right\rrbracket\left\llbracket\begin{array}{c}
i\omega a_{z}\\
i\omega\psi_{z}\\
i\omega h_{tz}
\end{array}\right\rrbracket^{-1}+\left\llbracket\begin{array}{c}
\mu h_{tz}^{\prime}\\
0\\
\mu a_{z}^{\prime}
\end{array}\right\rrbracket\left\llbracket\begin{array}{c}
i\omega a_{z}\\
i\omega\psi_{z}\\
i\omega h_{tz}
\end{array}\right\rrbracket^{-1}\nonumber\\
 &  & -\tilde{\tau}\left(\begin{array}{ccc}
-\frac{i\omega}{f} & 0 & 0\\
0 & -\frac{i\omega}{r^{2}f} & 0\\
0 & 0 & -\frac{i\omega}{r^{4}}
\end{array}\right)\tilde{\tau}\nonumber\\
 & = & \left(\begin{array}{ccc}
-\frac{i\omega}{f} & 0 & \frac{iq_{m}}{f}\\
0 & -\frac{i\omega r^{2}}{f} & -\frac{\beta r^{2}}{f}\\
-\frac{iq_{m}}{f} & -\frac{\beta r^{2}}{f} & -\frac{q_{m}^{2}+\beta^{2}r^{2}}{i\omega f}
\end{array}\right)+\left(\begin{array}{ccc}
0 & 0 & -\frac{\mu}{r^{4}}\\
0 & 0 & 0\\
-\frac{\mu}{f} & 0 & 0
\end{array}\right)\tilde{\tau}-\tilde{\tau}\left(\begin{array}{ccc}
-\frac{i\omega}{f} & 0 & 0\\
0 & -\frac{i\omega}{r^{2}f} & 0\\
0 & 0 & -\frac{i\omega}{r^{4}}
\end{array}\right)\tilde{\tau}.\label{eqflow}
\end{eqnarray}
Since equation (\ref{eq:Einstein}) is a first order differential equation,
the components of $\tilde{\tau}$ are not independent. We can recast equation (\ref{eq:Einstein}) as a constrained equation
\begin{eqnarray}
\left(\begin{array}{ccc}
0 & 0 & 0\\
iq_{m} & \beta & -i\omega\\
0 & 0 & 0
\end{array}\right)\tilde{\tau}
 & = & \left(\begin{array}{ccc}
0 & 0 & 0\\
-\mu & 0 & \frac{\mu q_{m}}{\omega}\\
0 & 0 & 0
\end{array}\right).
\end{eqnarray}
The constrained equations then take the form
\begin{equation}
\begin{cases}
iq_{m}\tilde{\tau}_{11}+\beta\tilde{\tau}_{21}-i\omega\tilde{\tau}_{31} & =-\mu,\\
iq_{m}\tilde{\tau}_{12}+\beta\tilde{\tau}_{22}-i\omega\tilde{\tau}_{32} & =0,\\
iq_{m}\tilde{\tau}_{13}+\beta\tilde{\tau}_{23}-i\omega\tilde{\tau}_{33} & =\frac{\mu q_{m}}{\omega}.
\end{cases}\label{eq:relations}
\end{equation}
%The near horizon analysis:
%\[
%f\tilde{\sigma}^{\prime}=\left(\begin{array}{ccc}
%-i\omega & 0 & iq_{m}\\
%0 & -i\omega r^{2} & -\beta r^{2}\\
%-iq_{m} & -\beta r^{2} & -\frac{q_{m}^{2}+\beta^{2}r^{2}}{i\omega}
%\end{array}\right)+\left(\begin{array}{ccc}
%0 & 0 & -\frac{\mu}{r^{4}}f\\
%0 & 0 & 0\\
%-\mu & 0 & 0
%\end{array}\right)\tilde{\sigma}-\tilde{\sigma}\left(\begin{array}{ccc}
%-i\omega & 0 & 0\\
%0 & -\frac{i\omega}{r^{2}} & 0\\
%0 & 0 & -\frac{i\omega}{r^{4}}f
%\end{array}\right)\tilde{\sigma}
%\]
%\[
%f=0\Longrightarrow(r=1)
%\]
Evaluating the flow equation (\ref{eqflow}) at the event horizon, we have
\[
0=\left(\begin{array}{ccc}
-i\omega & 0 & iq_{m}\\
0 & -i\omega\rh^2 & -\beta\rh^2\\
-iq_{m} & -\beta\rh^2 & -\frac{q_{m}^{2}+\beta^{2}\rh^2}{i\omega}
\end{array}\right)+\left(\begin{array}{ccc}
0 & 0 & 0\\
0 & 0 & 0\\
-\mu & 0 & 0
\end{array}\right)\tilde{\tau}_\mathrm{H}-\tilde{\tau}_\mathrm{H}\left(\begin{array}{ccc}
-i\omega & 0 & 0\\
0 & -i\omega/\rh^2 & 0\\
0 & 0 & 0
\end{array}\right)\tilde{\tau}_\mathrm{H}.
\]
Solving this equation, we achieve the horizon value of $\tilde{\tau}$
\bea
\tilde{\tau}_\mathrm{H}=\left(\begin{array}{ccc}
1 & 0 & -\frac{q_{m}}{\omega}\\
0 & \rh^2 & \frac{\beta\rh^2}{i\omega}\\
\frac{\mu+iq_{m}}{i\omega} & \frac{\beta\rh^2}{i\omega} &\tilde{\sigma}_{33}.
\end{array}\right).
\eea
%\[
%\tilde{\sigma}_{11}^{2}+\tilde{\sigma}_{12}\tilde{\sigma}_{21}=1
%\]
%\[
%\tilde{\sigma}_{11}\tilde{\sigma}_{12}+\tilde{\sigma}_{12}\tilde{\sigma}_{22}=0\Longrightarrow\tilde{\sigma}_{12}=0
%\]
%\[
%\Longrightarrow\tilde{\sigma}_{11}=1
%\]
%\[
%\tilde{\sigma}_{11}\tilde{\sigma}_{13}+\tilde{\sigma}_{12}\tilde{\sigma}_{23}=-\frac{q_{m}}{\omega}
%\]
%\[
%\Longrightarrow\tilde{\sigma}_{13}=-\frac{q_{m}}{\omega}
%\]
%\[
%\tilde{\sigma}_{21}\tilde{\sigma}_{11}+\tilde{\sigma}_{22}\tilde{\sigma}_{21}=0\Longrightarrow\tilde{\sigma}_{21}=0
%\]
%\[
%-\mu\tilde{\sigma}_{11}-\tilde{\sigma}_{31}(-i\omega)\tilde{\sigma}_{11}-\tilde{\sigma}_{32}(-i\omega)\tilde{\sigma}_{21}=iq_{m}
%\]
%\[
%\Longrightarrow\tilde{\sigma}_{31}=\frac{\mu+iq_{m}}{i\omega}
%\]
%\[
%\tilde{\sigma}_{21}\tilde{\sigma}_{12}+\tilde{\sigma}_{22}^{2}=1\Longrightarrow\tilde{\sigma}_{22}=1
%\]
%\[
%-\tilde{\sigma}_{21}(-i\omega)\tilde{\sigma}_{13}-\tilde{\sigma}_{22}(-i\omega)\tilde{\sigma}_{23}=\beta
%\]
%\[
%\Longrightarrow\tilde{\sigma}_{23}=\frac{\beta}{i\omega}
%\]
%\[
%-\mu\tilde{\sigma}_{12}-\tilde{\sigma}_{31}(-i\omega)\tilde{\sigma}_{12}-\tilde{\sigma}_{32}(-i\omega)\tilde{\sigma}_{22}=\beta
%\]
%\[
%\Longrightarrow\tilde{\sigma}_{32}=\frac{\beta}{i\omega}
%\]
%\[
%-\mu\tilde{\sigma}_{13}-\tilde{\sigma}_{31}(-i\omega)\tilde{\sigma}_{13}-\tilde{\sigma}_{32}(-i\omega)\tilde{\sigma}_{23}=\frac{q_{m}^{2}+\beta^{2}}{i\omega}
%\]
Note that $\tilde{\tau}_{33}$ cannot be derived from the flow equation. But from
the third line of (\ref{eq:relations}), we obtain
\[
\tilde{\tau}_{33}=\frac{1}{i\omega}(iq_{m}\tilde{\tau}_{13}+\beta\tilde{\tau}_{23}-\frac{\mu q_{m}}{\omega})=iq_{m}\frac{\mu+iq_{m}}{\omega^{2}}-\frac{\beta^{2}\rh^2}{\omega^{2}}.
\]
Finally we obtain the explicit form of $\tilde{\tau}_0$
\begin{eqnarray}
\tilde{\tau}_\mathrm{H}=\left(\begin{array}{ccc}
1 & 0 & -\frac{q_{m}}{\omega}\\
0 & \rh^2 & \frac{\beta\rh^2}{i\omega}\\
\frac{\mu+iq_{m}}{i\omega} & \frac{\beta\rh^2}{i\omega} & iq_{m}\frac{\mu+iq_{m}}{\omega^{2}}-\frac{\beta^{2}\rh^2}{\omega^{2}}
\end{array}\right).
\end{eqnarray}
As a consistency check, one can see that the first two lines of (\ref{eq:relations}) are
also satisfied by $\tilde{\tau}_\mathrm{H}$. Substituting the expression of $\tilde{\tau}_\mathrm{H}$ into (\ref{definesigma}), we obtain
\begin{eqnarray}
-f a_{z}^{\prime} & \to & i\omega a_{z}-iq_{m}h_{tz}\big|_{r=\rh},\label{azhorizon}\\
-f \psi_{z}^{\prime} & \to & i\omega\rh^2\psi_{z}+\beta\rh^2 h_{tz}\big|_{r=\rh},\\
-\rh^4 h'_{tz}& \to & (\mu+i q_m)a_z+\beta \rh^2 \psi_z-\frac{\mu q_m+i q^2_m}{\omega}h_{tz}-\frac{\beta^2 \rh^2}{\omega}h_{tz}\big|_{r=\rh}.
\end{eqnarray}
Imposing the horizon regularity condition on equation (\ref{htx}) and the relation given in (\ref{azhorizon}), we find that
\bea
(q^2_m+\beta^2 \rh^2)h_{tz}\bigg|_{r=\rh}&=&\mu f a'_z+\omega q_m a_z-i\beta \omega \rh^2 \psi_z\bigg|_{r=\rh}\nonumber\\
&=&-i\omega \mu a_z+i \mu q_m h_{tz}+\omega q_m a_z-i\beta \omega \rh^2 \psi_z\bigg|_{r=\rh}.
\eea
We then obtain
\bea
h_{tz}\bigg|_{r=\rh}&=&-i \omega a_z\frac{\mu+i q_m}{q^2_m+\beta^2 \rh^2-i q_m \mu}-\frac{i\beta \omega \rh^2 \psi_z}{q^2_m+\beta^2 \rh^2-i q_m \mu}\bigg|_{r=\rh}.
\eea
Bearing in mind the relation $h_{tz}=(h_{tx}-ih_{ty})/2$ and $\psi_z=(\psi_x-i\psi_y)/2$, we obtain the expression for $h_{tx}$ and $h_{ty}$
\bea
h_{tx}&=&-i\omega\frac{\mu\beta^2\rh^2 a_x+q_m(q^2_m+\beta^2\rh^2+\mu^2)a_y}{q^4_m+\beta^4 \rh^4+q^2_m(\mu^2+2\beta^2\rh^2)}-i\omega\frac{\beta \rh^2[(q^2_m+\beta^2\rh^2)\psi_x-\mu q_m \psi_y]}{q^4_m+\beta^4 \rh^4+q^2_m(\mu^2+2\beta^2\rh^2)},\label{exhtx}\\
h_{ty}&=&-i\omega\frac{\mu\beta^2\rh^2 a_y-q_m(q^2_m+\beta^2\rh^2+\mu^2)a_x}{q^4_m+\beta^4 \rh^4+q^2_m(\mu^2+2\beta^2\rh^2)}-i\omega\frac{\beta \rh^2[(q^2_m+\beta^2\rh^2)\psi_y-\mu q_m \psi_x]}{q^4_m+\beta^4 \rh^4+q^2_m(\mu^2+2\beta^2\rh^2)}.\label{exhty}
\eea
We can drop out the second term in (\ref{exhtx}) and (\ref{exhty}) in the following calculation since the $\psi_x$ and $\psi_y$ do not contribute to the transport coefficients.
The radially conserved currents can be deduced from the Maxwell equation
\bea
&&J_{x}=\sqrt{\frac{g_{tt}}{g_{rr}}}Z a'_x-\mu \rh h_{tx}=i\omega Z a_x-\mu \rh h_{tx}-Z B h_{ty},\nonumber\\
&&J_{y}=\sqrt{\frac{g_{tt}}{g_{rr}}}Z a'_y-\mu \rh h_{ty}=i\omega Z a_y-\mu \rh h_{ty}+Z B h_{tx}.
\eea
The electric conductivity tensor can be evaluated via $\sigma_{ij}\equiv \partial J_i/\partial E_j$  with $E_j=i\omega a_j$, that is to say
\bea
&&\sigma_{xx}=\sigma_{yy}=\beta^2 \rh^2 \frac{B^2+\rh^2(\mu^2+\beta^2)}{\rh^2\mu^2B^2+(B^2+\beta^2\rh^2)^2},\\
&&\sigma_{xy}=-\sigma_{yx}=B\mu \rh \frac{B^2+\rh^2(\mu^2+\beta^2)}{\rh^2\mu^2B^2+(B^2+\beta^2\rh^2)^2}.
\eea
The radially conserved heat current still takes the form (\ref{heatcurrent}), i.e. $\mathcal{Q}_i=-4\pi T g_{xx}h_{tx_i}$.
The thermoelectric conductivity tensor calculated at the event horizon is then given by
\bea
&&\alpha_{xx}=\alpha_{yy}=\frac{4\pi \rh^5 \beta^2 \mu}{B^2 \mu^2 \rh^2+(B^2+\beta^2\rh^2)^2},\\
&&\alpha_{xy}=-\alpha_{yx}=4\pi B \rh^3 \frac{B^2+\rh^2(\mu^2+\beta^2)}{B^2 \mu^2 \rh^2+(B^2+\beta^2\rh^2)^2}.
\eea
Having obtained $\sigma_{ij}$ and $\alpha_{ij}$, one can utilize the equations below to calculate the thermal conductivity
\begin{equation}
\label{pheno}
\left(\begin{array}{c}\langle J_{i} \rangle \\ \langle Q_{i} \rangle \end{array}\right)
=
\left(\begin{array}{cc}   {\sigma}_{ij} &{\alpha}_{ij} T \\  { \bar{\alpha}}_{ij} T & {\bar{\kappa}}_{ij} T \end{array}\right)
\left(\begin{array}{c} E_{j} \\ - (\nabla_{j} T)/T\end{array}\right)~,
\end{equation}
Considered the condition $Q_x=0$, $\nabla_{y} T=0$ and $E_y=0$, the thermal conductivity matrix component $\bar{\kappa}_{xx}$ is given by
\bea
\bar{\kappa}_{xx}=\frac{T \alpha^2_{xx}}{\sigma_{xx}-\sigma^0_{xx}},
\eea
where $\sigma^0_{xx}$ is the electric conductivity with vanishing heat current that is considered as an input \cite{gsw2015}.
Similarly, the case $Q_x=0$, $\nabla_{x} T=0$ and $E_y=0$ gives
\be
\bar{\kappa}_{xy}=\frac{T \alpha_{xx}\alpha_{xy}}{\sigma_{xx}}.
\ee
Therefore, we finally obtain
\bea
&&\bar{\kappa}_{xx}=\bar{\kappa}_{yy}=\frac{16\pi^2 T \rh^4(B^2+\beta^2 \rh^2)}{B^2 \mu^2 \rh^2+(B^2+\beta^2\rh^2)^2},\\
&&\bar{\kappa}_{xy}=-\bar{\kappa}_{yx}=\frac{16\pi^2 T B \mu\rh^5}{B^2 \mu^2 \rh^2+(B^2+\beta^2\rh^2)^2}.
\eea
The above expressions exactly reproduce the results given in \cite{kim2,mb3,mb4}. Carrying on the numerical computation, we can also reproduce the optical conductivity same as \cite{kim2}. In all, our method works both for non-magnetic and magnetic systems.

\section{Conclusions}
In summary, we developed a new approach for the calculation of transport coefficients in holographic systems, particularly with translational symmetry breaking. We first started from Onsager's theory and generalized the {holographic flow of transport coefficients to the case of coupled transportation}. As a concrete example, we revisited the RN-AdS black hole with linear axion fields and calculate the transport coefficients analytically and numerically. Since the main equation is a first-order ordinary differential equation, the computation here is somehow simplified compared to the method given in \cite{kim}.

The advantages of this method can be summarized as follows. Firstly, it works for coupled flow equations of transport. In \cite{sin}, the authors have not obtained the explicit flow equations when facing with coupled equations consisted of different modes of perturbations. We have shown that by using the equation of motion repeatedly, we can finally obtain the flow equation in an elegant matrix form, which can have the direct physical meaning of the RG flow equation of the transport coefficients. Secondly, the numerical calculation of the AC conductivity could be greatly simplified by just solving the first-order nonlinear ordinary differential equation. Thirdly, our approach can also be easily utilized for coupled linear transport with finite momentum $k$, which provides a powerful tool (at least) for numerical study of such kind of physics. For spatially inhomogeneous systems, our method with these advantages should be particularly efficient, since it only needs a simple Runge-Kutta marching instead of, say, the resource-consuming pseudo-spectral method. We expect to extend our approach to those systems.

Finally, we notice that \cite{ayan11, ayan13} work on the holographic RG flow of the transport coefficients without momentum dissipation in the hydrodynamic regime. It would be interesting to compare our work with theirs. But it is beyond the scope of this paper. It would also be interesting to explore the possible correlation between holography and the optimization theory (or other disciplines), since the same type of matrix Riccati equation appears.

\section*{Acknowledgement}
We would like to thank Elias Kiritsis, Hong L$\rm\ddot{u}$, Sang-Jin Sin and John McGreevy for helpful discussions.
YT is partially supported by NSFC with Grant No.11475179 and the Grant (No. 14DZ2260700) from Shanghai Key Laboratory of High Temperature Superconductors. He is also partially supported by the ``Strategic Priority Research Program of the Chinese Academy of Sciences", Grant No. XDB23030000. XHG was partially supported by NSFC,
China (No.11375110). SFW was supported partially by NSFC China (No. 11675097).

\appendix
\section{{The flow equation in the Einstein-Maxwell theory}}
The linear perturbations on top of the $d+1$ dimensional RN-AdS bulk spacetime have the following effective action \cite{sin2}:
\begin{equation}\label{effective}
I=\int d^{d+1}x\sqrt{-g}\left(-\frac{1}{4}F_{\mu\nu}F^{\mu\nu}-\frac{1}{4e^{2}}f_{\mu\nu}f^{\mu\nu}+a_{t}A_{x}^{\prime}\bar{A}_{t}^{\prime}-a_{r}\partial_{t}A_{x}\bar{A}_{t}^{\prime}\right),
\end{equation}
where $a_{\mu}=h_{\mu}^{x}$ is the vector mode of metric perturbations with the effective coupling
\begin{equation}\label{}
e^2\equiv\frac{16\pi G}{g_{xx}}
\end{equation}
and
\begin{equation}\label{}
f_{\mu\nu}=\partial_{\mu}a_{\nu}-\partial_{\nu}a_{\mu}.
\end{equation}
Hereafter, we take the radial gauge $a_r=0$ for convenience. We consider the case that all the perturbations have no momentum, which means that the coupled perturbations are
\begin{equation}\label{}
{\Phi}=\left(\begin{array}{c}
A_x\\
a_t
\end{array}\right)
\end{equation}
and the effective action (\ref{effective}) becomes
\begin{equation}\label{}
I=\int d^{d+1}x\sqrt{-g}\left( -\frac{1}{2}g^{rr}g^{xx}A_{x}^{\prime2}-\frac{1}{2}g^{tt}g^{xx}\dot{A}_{x}^{2}-\frac{1}{2e^{2}}g^{rr}g^{tt}a_{t}^{\prime2}+a_{t}A_{x}^{\prime}\bar{A}_{t}^{\prime} \right).
\end{equation}
{Here an over dot means derivative with respect to $t$.} So we have
\begin{equation}\label{}
\left(\begin{array}{c}
J^x\\
j^t
\end{array}\right)=\frac{{\delta} L}{{\delta}\Phi^{\prime}}=\sqrt{-g}\left( \begin{array}{c}
-g^{rr}g^{xx}A_{x}^{\prime}+a_t\bar{A}_{t}^{\prime}\\
-e^{-2}g^{rr}g^{tt}a_{t}^{\prime}
\end{array} \right),
\end{equation}
and
\begin{equation}\label{}
\frac{{\delta} L}{{\delta}\Phi}=\sqrt{-g}\left(\begin{array}{c}
	-\omega^{2}g^{tt}g^{xx}A_{x}\\
	A_{x}^{\prime}\bar{A}_{t}^{\prime}
\end{array}\right).
\end{equation}
Then the transport matrix has the form
\begin{eqnarray}\label{}
\tau&=&\sqrt{-g}\left\llbracket\begin{array}{c}
	-g^{rr}g^{xx}A_{x}^{\prime}+a_{t}\bar{A}_{t}^{\prime}\\
	-e^{-2}g^{rr}g^{tt}a_{t}^{\prime}
\end{array}\right\rrbracket\left\llbracket\begin{array}{c}
	i\omega A_{x}\\
	i\omega a_{t}
\end{array}\right\rrbracket^{-1}\nonumber\\
&=&\sqrt{-g}\left\llbracket\begin{array}{c}
-g^{rr}g^{xx}A_{x}^{\prime}\\
-e^{-2}g^{rr}g^{tt}a_{t}^{\prime}
\end{array}\right\rrbracket\left\llbracket\begin{array}{c}
i\omega A_{x}\\
i\omega a_{t}
\end{array}\right\rrbracket^{-1}+\frac{\sqrt{-g}}{i\omega}\left(\begin{array}{cc}
0 & \bar{A}_{t}^{\prime} \\
0& 0
\end{array}\right).
\end{eqnarray}
The flow equation (\ref{eq:raw}) becomes
\begin{eqnarray}
\tau^{\prime}&=&\sqrt{-g}\left\llbracket\begin{array}{c}
-\omega^{2}g^{tt}g^{xx}A_{x}\\
A_{x}^{\prime}\bar{A}_{t}^{\prime}
\end{array}\right\rrbracket \left\llbracket\begin{array}{c}
i\omega A_{x}\\
i\omega a_{t}
\end{array}\right\rrbracket^{-1}-\tau\left\llbracket\begin{array}{c}
A_x^{\prime}\\
a_t^{\prime}
\end{array}\right\rrbracket \left\llbracket\begin{array}{c}
A_x\\
a_t
\end{array}\right\rrbracket^{-1}\nonumber\\
&=&\sqrt{-g}\left(\begin{array}{cc}
i\omega g^{tt}g^{xx} & 0\\
0 & \frac{\bar{A}_{t}^{\prime 2}}{i\omega g^{rr}g^{xx}}
\end{array}\right)-\left(\begin{array}{cc}
0 & 0\\
\frac{\bar{A}_{t}^{\prime}}{g^{rr}g^{xx}} & 0
\end{array}\right)\tau-\tau\left(\begin{array}{cc}
0 & \frac{\bar{A}_{t}^{\prime}}{g^{rr}g^{xx}}\\
0 & 0
\end{array}\right)\nonumber\\
&&-\frac{i\omega}{\sqrt{-g}}\tau\left(\begin{array}{cc}
-\frac{1}{g^{rr}g^{xx}} & 0\\
0 & -\frac{e^{2}}{g^{rr}g^{tt}}
\end{array}\right)\tau.\label{eq:flow}
\end{eqnarray}
It can be seen that from this construction the flow equation is always invariant under the transposition of $\tau$, which obviously reflects Onsager's reciprocal relation. The above flow equation takes almost the same form of the matrix Riccati equation as in optimization theory and other disciplines (see, for example, \cite{optimization}), where the derivation is with respect to the time $t$ instead of the holographic direction $r$.

In all the cases considered here, the horizon is located at
\begin{equation}\label{}
g_{tt}\sim g^{rr}\to 0,
\end{equation}
so multiplying both sides of (\ref{eq:flow}) by $g_{tt}$ gives rise to
\begin{eqnarray}
0&=&\sqrt{-g}\left(\begin{array}{cc}
i\omega g^{xx} & 0\\
0 & \frac{g_{tt}\bar{A}_{t}^{\prime 2}}{i\omega g^{rr}g^{xx}}
\end{array}\right)-\left(\begin{array}{cc}
0 & 0\\
\frac{g_{tt}\bar{A}_{t}^{\prime}}{g^{rr}g^{xx}} & 0
\end{array}\right)\tau_\mathrm{H}-\tau_\mathrm{H}\left(\begin{array}{cc}
0 & \frac{g_{tt}\bar{A}_{t}^{\prime}}{g^{rr}g^{xx}}\\
0 & 0
\end{array}\right)\nonumber\\
&&-\frac{i\omega}{\sqrt{-g}}\tau_\mathrm{H}\left(\begin{array}{cc}
-\frac{g_{tt}}{g^{rr}g^{xx}} & 0\\
0 & 0
\end{array}\right)\tau_\mathrm{H},
\end{eqnarray}
where $\tau_\mathrm{H}$ is the horizon value of $\tau$. In the RN-AdS case, $g_{tt}=-g^{rr}$, so we have
\begin{eqnarray}\label{}
0&=&\sqrt{-g}\left(\begin{array}{cc}
i\omega g^{xx} & 0\\
0 & -\frac{\bar{A}_{t}^{\prime 2}}{i\omega g^{xx}}
\end{array}\right)-\left(\begin{array}{cc}
0 & 0\\
-\frac{\bar{A}_{t}^{\prime}}{g^{xx}} & 0
\end{array}\right)\tau_\mathrm{H}-\tau_\mathrm{H}\left(\begin{array}{cc}
0 & -\frac{\bar{A}_{t}^{\prime}}{g^{xx}}\\
0 & 0
\end{array}\right)\nonumber\\
&&-\frac{i\omega}{\sqrt{-g}}\tau_\mathrm{H}\left(\begin{array}{cc}
\frac{1}{g^{xx}} & 0\\
0 & 0
\end{array}\right)\tau_\mathrm{H}.
\end{eqnarray}
The above matrix equation can be solved as
\begin{equation}\label{}
\tau_\mathrm{H}=\sqrt{-g}\left(\begin{array}{cc}
g^{xx} & \frac{\bar{A}_{t}^{\prime}}{i\omega}\\
\frac{\bar{A}_{t}^{\prime}}{i\omega} & 0
\end{array}\right).
\end{equation}
Using this $\tau_\mathrm{H}$ as the IR input of the flow equation (\ref{eq:flow}), we can obtain $\tau(r)$ as a symmetric matrix numerically for nonzero $\omega$ and analytically in the $\omega\to 0$ limit. Actually, the 1-1 component of $\tau(r\to\infty)$ is just the (AC or DC) electric conductivity of the boundary system, while the thermal conductivity and thermoelectric conductivity are linear combinations of the components of $\tau$ (see for example \cite{review}).

\end{document}